# Spin Hall effect in 3d ferromagnetic metals for field-free switching of perpendicular magnetization: A first-principles investigation

Fanxing Zheng[†], Jianting Dong[†], Yizhuo Song, Meng Zhu, Xinlu Li, and Jia Zhang[*]

[1]*School of Physics and Wuhan National High Magnetic Field Center, Huazhong University of Science and Technology, 430074 Wuhan, China.*

[*]jiazhang@hust.edu.cn
[†]Authors contributed equally to this work.

## ABSTRCT

Ferromagnetic metals, with the potential to generate spin current with unconventional spin polarization via the spin Hall effect, offer promising opportunities for field-free switching of perpendicular magnetization and for the spin-orbit torque devices. In this study, we investigate two distinct spin Hall mechanisms in 3d ferromagnetic metals including spin-orbit coupling driven spin Hall effect in Fe, Co, Ni and their alloys, and non-relativistic spin Hall effect arising from anisotropic spin-polarized transport by taking $L1_0$-MnAl as an example. By employing first-principles calculations, we examine the temperature and alloy composition dependence of spin Hall conductivity in Fe, Co, Ni and their alloys. Our results reveal that the spin Hall conductivities with out-of-plane spin polarization in 3d ferromagnetic metals are at the order of 1000 $\hbar/2e(\Omega cm)^{-1}$ at 300 K, but with a relatively low spin Hall angles around 0.01~0.02 due to the large longitudinal conductivity. For $L1_0$-MnAl(101), the non-relativistic spin Hall conductivity can reach up to 10000 $\hbar/2e(\Omega cm)^{-1}$, with a giant spin Hall angle around 0.25 at room temperature. By analyzing the magnetization switching process, we demonstrate deterministic switching of perpendicular magnetization without an external magnetic field by using 3d ferromagnetic metals as spin current sources. Our work may provide an unambiguous understanding on spin Hall effect in ferromagnetic metals and pave the way for their potential applications in related spintronic devices.

# INTRODUCTION

Spintronics excavates the potential of incorporating spin degrees of freedom into traditional charge-based devices or entirely replacing charge with spin functionality, which provides a promising pathway for designing advanced spintronic devices [1-2]. Various schemes have been proposed to control spin via electrical methods [3-5]. Due to the ubiquity of spin-orbit coupling (SOC), current approaches primarily utilize spin-orbit torque (SOT) generated by the spin Hall effect to manipulate magnetization in various applications [6]. One of the most important advantages of the SOT switching scheme is the separation of writing and reading path in SOT-based magnetic random-access memory (MRAM) [7]. Previous research on SOT switching has employed non-magnetic heavy metals Pt, Ta, W etc. as spin current sources in "heavy metal/ferromagnet" bilayers [8-15]. In such case, an additional external in-plane magnetic field is required to switch perpendicular magnetization, which is challenging for device integration.

Instead of using non-magnetic materials, an efficient approach to switching perpendicular magnetization may involve utilizing ferromagnetic metals as spin current sources [16-23]. Several theoretical and experimental studies have demonstrated that the SOC-driven magnetic spin Hall effect in ferromagnetic materials may generate unconventional out-of-plane spin polarization, enabling deterministic field-free switching of perpendicular magnetization [24-28]. Meanwhile, recent theory predicts a significant non-relativistic spin Hall mechanism in magnetic metals with anisotropic transport spin polarization [29]. Therefore, comprehensive understanding on spin Hall effect in ferromagnetic metals is urgently needed to identify practical SOT applications involving ferromagnetic materials and the efficiency for achieving field-free switching.

In this work, we investigate the spin Hall effect in 3$d$ ferromagnetic metals including spin Hall effect driven by spin-orbit coupling in Fe, Co, Ni and their alloys, as well as non-relativistic spin Hall effect in L1$_0$-MnAl by using first-principles calculations. In addition, we numerically solve the Landau-Lifshitz-Gilbert (LLG) equation by adopting the calculated spin Hall conductivity and angle to analyze the switching efficiency of perpendicular magnetization.

## COMPUTATIONAL METHOD

The first-principles calculations are performed by using the multiple-scattering Korringa–Kohn–Rostoker Green's function (KKR-GF) method [30-31]. A cutoff $l_{max}$ =3 was employed for the angular momentum expansion, and the Vosko-Wilk-Nussair (VWN) type of local density approximation (LDA) was utilized for exchange-correction potential for Fe, Co, Ni and their alloys [32], while Perdew-Burke-Ernzerhof (PBE) was applied for $L1_0$-MnAl [33]. For self-consistent calculations, energy integration was performed on a semicircle in the complex plane by using 30 energy points and 45×45×45 $k$-points in the Brillouin zone. After achieving self-consistent potential, the conductivities were calculated within Kubo-Bastin linear response formalism [34]. Approximately $10^7$ $k$-points in the Brillouin zone were found to be sufficient for the convergence of spin Hall conductivity. The Coherent Potential Approximation was employed to deal with chemical and thermally induced disorder scattering [35-38]. For SOC-driven spin Hall effect in pure metals Fe, Co, Ni and the alloys $Fe_xCo_{1-x}$ and $Ni_xFe_{1-x}$ ($0.3 \leq x \leq 0.7$), experimental lattice parameters with various concentrations were adopted in this work [39-41]. Additionally, the Bloch spectra function was utilized to visualize the temperature dependent electronic structure of Fe, representing the spectral weight distribution of electrons in momentum space [42].

## RESULTS AND DISCUSSIONS

### A. Symmetry-imposed spin current in ferromagnetic metals

In linear-response theory, the spin current propagates along the $\mu$ axis with spin polarization direction $s$ by applying an electric field in the $\nu$ axis, which can be described as

$$J_\mu^s = \sum_\nu J_{\mu\nu}^s = \sum_\nu \sigma_{\mu\nu}^s E_\nu \quad (\mu, \nu, s \in \{x, y, z\}), \tag{1}$$

where the spin Hall conductivity $\sigma_{\mu\nu}^s$ is a third-rank tensor. In this section, we focus on crystals with cubic symmetry, which possess inversion symmetry, two-fold rotational symmetries around the $x$, $y$, and $z$ axes, as well as mirror symmetries $M_z$, $M_y$, and $M_x$. As illustrated in Fig. 1(a), when a charge current is applied along $x$ axis, for non-magnetic materials it exhibits $M_y$ and $M_z$ mirror symmetries and two-fold rotational symmetry around $x$ axis ($C_{2x}$). Consequently, the spin current generated along the $z$ axis by the spin Hall effect can only contain $S_y$ spin polarization. In the case of a ferromagnetic metal, the presence of magnetization reduces symmetries, allowing for spin currents with unconventional spin polarization. For example, as illustrated in Fig. 1(b), when the magnetization is aligned parallel to the charge current, only the two-fold rotational symmetry around the $x$ axis remains, while the $M_y$ and $M_z$ mirror symmetries are broken. As a result, the spin current along the $z$ axis may exhibit both $S_y$ and $S_z$ spin polarization. In contrast, when the magnetization is perpendicular to the applied current, as shown in Fig. 1(c), the generated spin current has only $S_y$ spin polarization, due to the presence of $M_y$ mirror symmetry. Breaking $M_y$ symmetry is therefore essential for generating $S_z$ spin current.

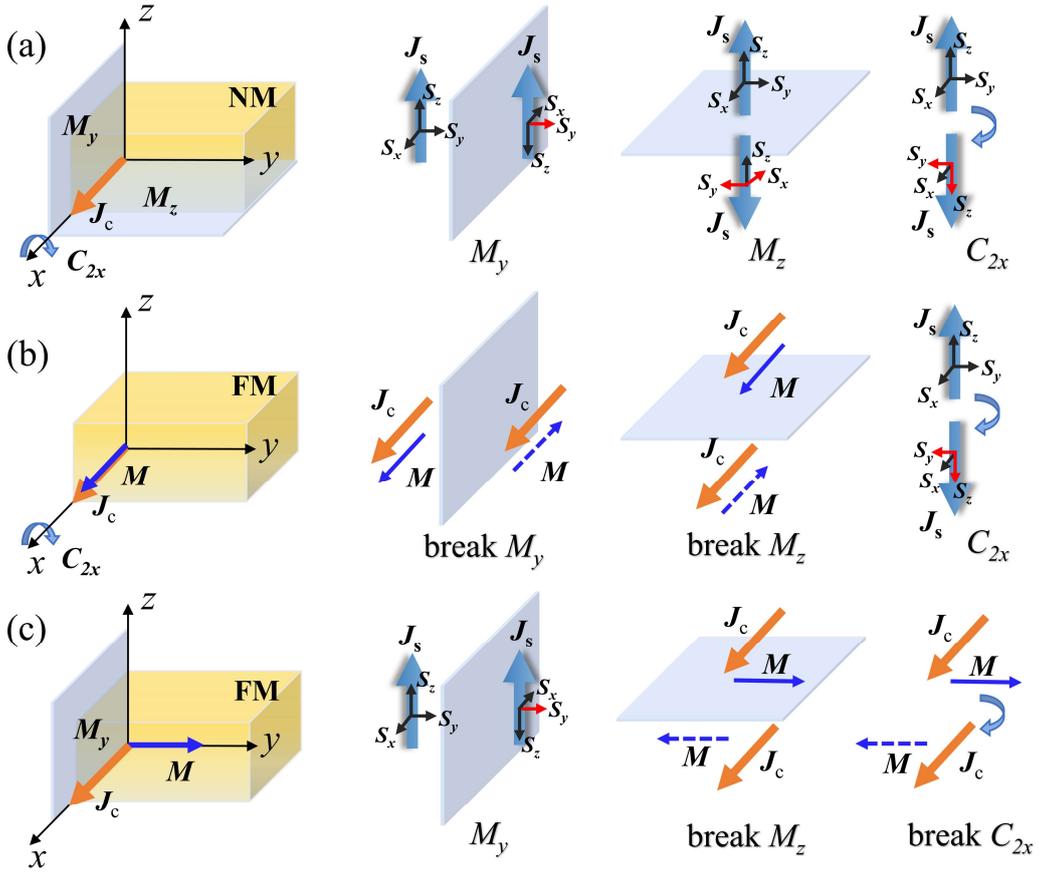

FIG. 1. Illustration of typical symmetry and spin current generated by spin Hall effect in nonmagnets (NM) and ferromagnets (FM). (a) In NM like fcc-Pt, bcc-Ta, W with charge current flowing along $x$ axis, spin current propagate along $z$ axis with only $S_y$ spin polarization is symmetry allowed due to the mirror planes, i.e., $M_y$ and $M_z$ as well as two-fold rotational

symmetry around *x* axis ($C_{2x}$). (b) In FM, when the magnetization *M* is parallel to the applied current, spin current along *z* axis generated by spin Hall effect may contain $S_y$ and $S_z$ spin polarization due to breaking $M_y$ and $M_z$ mirror symmetries. (c) When the magnetization *M* is perpendicular to the applied current, the spin current along *z* axis contains only $S_y$ spin polarization due to the presence of $M_y$ symmetry.

Based on the symmetry analysis, the SOC-driven spin Hall effect in 3*d* magnetic metals and alloys are investigated by setting magnetization parallel to electric field direction. In this situation, the spin current propagating along *z* axis exhibits $S_y$ and $S_z$ spin polarization, corresponding to the spin Hall conductivity elements of $\sigma_{zx}^{y}$ and $\sigma_{zx}^{z}$. Notably, such spin current may provide an opportunity for deterministic switching of perpendicular magnetization.

## B. Spin Hall effect in Fe, Co, Ni, and their alloys

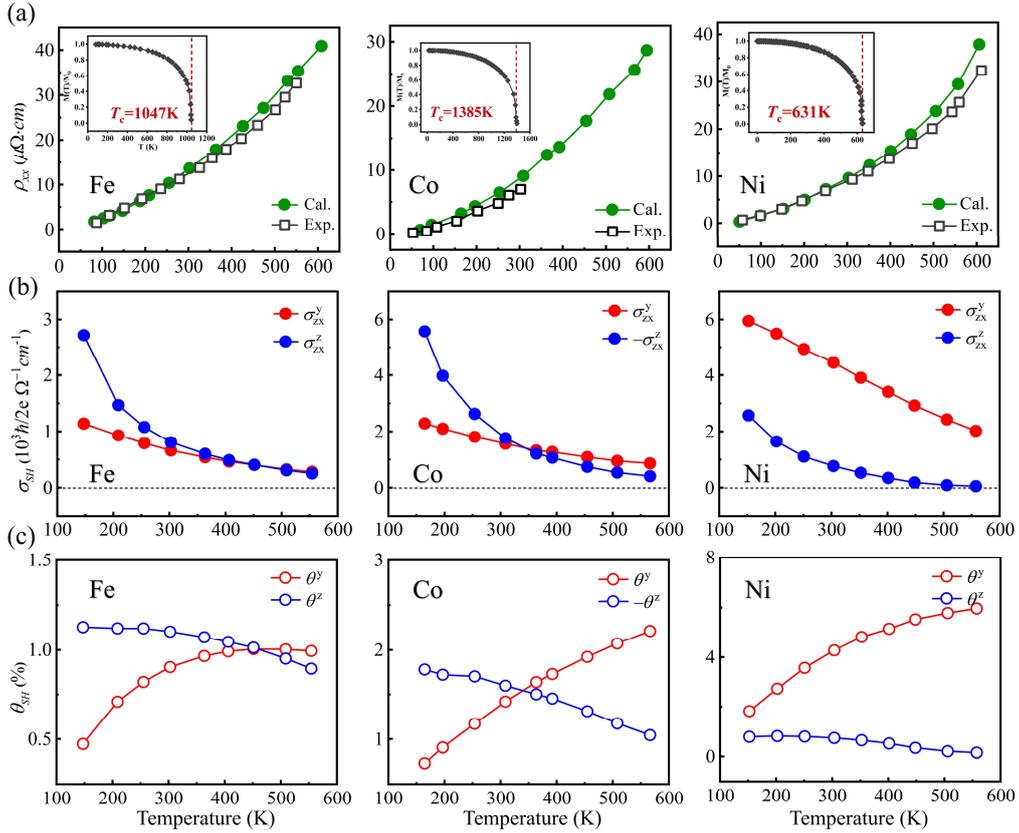

FIG. 2. The calculated temperature dependent (a) longitudinal resistivity $\rho_{xx}$ (Green circle) in comparison with its available experimental values (Black open square) [41][43]. The insets are experimental magnetization-temperature curves used for simulating spin disorder scattering. (b) The temperature dependent spin Hall conductivity $\sigma_{zx}^{y}$, $\sigma_{zx}^{z}$, and (c) the corresponding spin Hall angle $\theta^{y}$ and $\theta^{z}$ in Fe, Co, and Ni.

We first investigated the spin Hall effect driven by SOC in conventional Fe, Co, Ni and their alloys through full relativistic transport calculations. Fig. 2 shows the temperature dependence of spin Hall conductivity in a wide temperature range (50 < T < 500 K), with both phonon and spin disorder scattering taken into account. As shown in Fig. 2(a), the calculated electrical resistivity agrees well with experimental values [41][43], validating the accuracy of our calculations. Fig. 2(b) presents the representative spin Hall conductivity $\sigma_{zx}^{y}$ and $\sigma_{zx}^{z}$ as a function of temperature. It can be observed that both spin Hall conductivities decrease with increasing temperature, owing to enhanced phonon and spin disorder scattering, as well as band energy broadening (see the discussion in Appendix A). At room temperature (300 K), the calculated spin Hall conductivity $\sigma_{zx}^{z}$ with $S_z$ spin polarization is found to be 806 $\hbar/2e(\Omega cm)^{-1}$, -1758 $\hbar/2e(\Omega cm)^{-1}$, and 768 $\hbar/2e(\Omega cm)^{-1}$ in Fe, Co, and Ni, respectively. These values are significantly higher than other reported materials, such as $WTe_2/PtTe_2$ heterostructure (250 $\hbar/2e(\Omega cm)^{-1}$) [44], non-collinear antiferromagnet $Mn_3Pt$ (275 $\hbar/2e(\Omega cm)^{-1}$) and $Mn_3Ir$ (143 $\hbar/2e(\Omega cm)^{-1}$) [45-46]. Additionally, the spin Hall conductivity $\sigma_{zx}^{z}$ gradually vanishes as the temperature approaches Curie temperature ($T_c$), as clearly seen in the case of Ni ($T_c$ = 631K), confirming its magnetic origin. The corresponding spin Hall angles are defined as $\theta^{y,z} = (2e/\hbar)\sigma_{zx}^{y,z}/\sigma_{xx}$ and shown in Fig. 2(c). It can be seen that the $\theta^z$ is around 0.01 for Fe, Ni and 0.02 for Co below room temperature. As the temperature increases, the magnetization decreases until it vanishes at $T_c$, leading to a reduction of $\theta^z$ at high temperatures. In contrast, the $\theta^y$ increases by approximately a factor of 2 with increasing temperature for Fe, Co, and Ni.

Taking Fe as an example, the scaling relation between spin Hall conductivity with longitude conductivity $\sigma_{xx}$ has been investigated. As shown in Fig. 3, the spin Hall conductivity $\sigma_{zx}^{z}$ is found to depend linearly on the magnetization $M(T)$ and $\sigma_{xx}$, which can be well scaled by $\sigma_{zx}^{z} \sim M(T)/M_0 \cdot \sigma_{xx}$. Such scaling behavior is universal for other ferromagnetic materials, since both the $T^{odd}$ spin Hall conductivity $\sigma_{zx}^{z}$ (change sign under time reversal operation) and the longitudinal conductivity $\sigma_{xx}$ exhibit a similar dependence on band energy broadening $\hbar\Gamma$. Furthermore, the spin Hall conductivity $\sigma_{zx}^{y}$ follows a quadratic scaling with $\sigma_{xx}$, which can be well described as $\sigma_{zx}^{y} \sim a\sigma_{xx}^{2}$. In the clean limit $\Gamma \rightarrow 0$ (*i.e.*, in the higher conductivity region), the $T^{even}$ spin Hall conductivity $\sigma_{zx}^{y}$ (sign does not change under time reversal operation) converges to a constant value, which leads to small spin Hall angle $\theta^{y}$ at low temperatures and gradually increases at high temperatures.

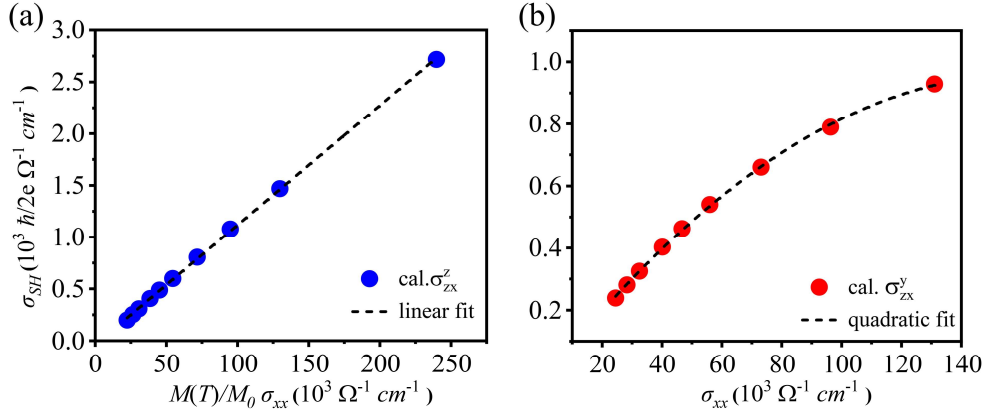

FIG. 3. The calculated spin Hall conductivity $\sigma_{zx}^{z}$ and $\sigma_{zx}^{y}$ of Fe as a function of longitudinal conductivity $\sigma_{xx}$. The dashed lines represent linear fitting by $\sigma_{zx}^{z} \sim M(T)/M_0 \cdot \sigma_{xx}$ in (a) and quadratic fitting by $\sigma_{zx}^{y} \sim a\sigma_{xx}^{2}$ in (b).

The composition dependence of the spin Hall effect in 3$d$ ferromagnetic alloys Fe$_x$Co$_{100-x}$ and Ni$_{100-x}$Fe$_x$ ($0.3 \leq x \leq 0.7$) is further studied. As shown in Fig. 4(a), the spin Hall conductivities $\sigma_{zx}^{y}$ and $\sigma_{zx}^{z}$ are all at the order of 1000 $\hbar/2e(\Omega cm)^{-1}$ for both alloys. Notably, the magnetic spin Hall conductivity $\sigma_{zx}^{z}$ in Fe$_x$Co$_{100-x}$ is more pronounced than that in the Ni$_{100-x}$Fe$_x$ alloy, which may be attributed to the larger magnetic moments in Fe$_x$Co$_{100-x}$ alloy (see Appendix B for details). At low Fe concentrations, the magnetization-independent $\sigma_{zx}^{y}$ dominates in Ni$_{100-x}$Fe$_x$ alloys. The corresponding spin Hall angles for Fe$_x$Co$_{100-x}$ and Ni$_{100-x}$Fe$_x$ are presented in Fig. 4(b). For both alloys, the spin Hall angle $\theta^{z}$, necessary for field-free switching of perpendicular magnetization, ranges between 0.01 and 0.02 and shows a non-obvious dependence on alloy composition.

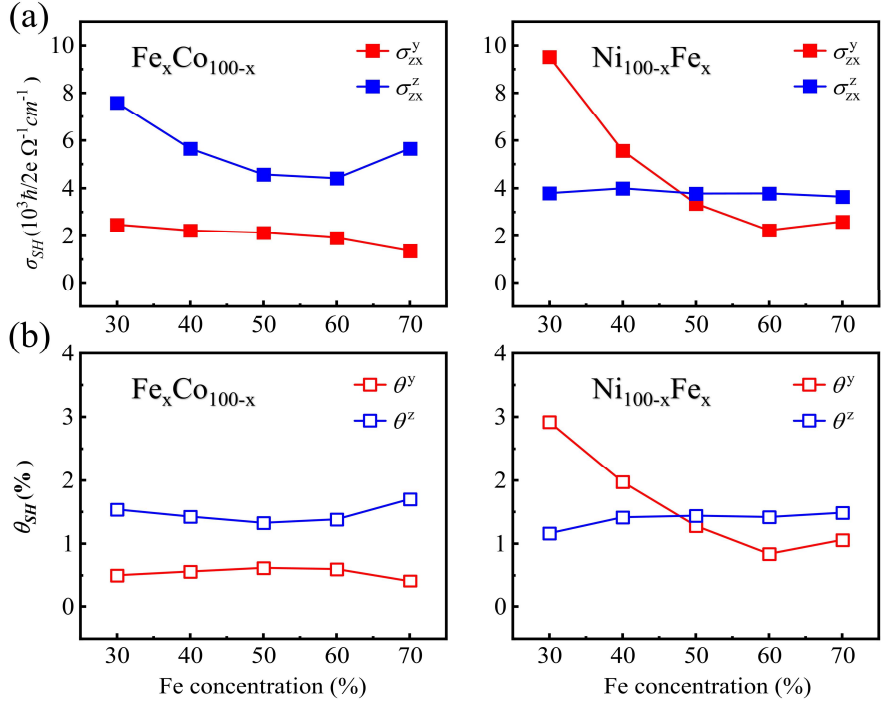

FIG. 4. The calculated composition dependent (a) spin Hall conductivity $\sigma_{zx}^{y}$, $\sigma_{zx}^{z}$ and (b) their corresponding spin Hall angle $\theta^{y}$ and $\theta^{z}$ in $Fe_xCo_{100-x}$ and $Ni_{100-x}Fe_x$ alloys, respectively.

### C. Non-relativistic spin Hall effect in L1$_0$-MnAl

In ferromagnets with lower symmetry, the existence of large anisotropic longitudinal spin polarization may generate remarkable non-relativistic spin Hall conductivity [29]. As a demonstration, we select L1$_0$-MnAl which has a magnetic moment of 2.5 $\mu_B$ per formula cell along the $c$ axis and Curie temperature above 600 K. Its lattice constant is $a=b=2.76$ Å, $c=3.48$ Å in the reduced tetragonal unit cell [47]. The structure of tetragonal MnAl(001) is shown in Fig. 5(a) with $c$ along $z$ axis, $a$, $b$ along $x$ and $y$ axes. Tetragonal MnAl belongs to the magnetic space group of $P_4/mm'm'$, which implies at least three mirror planes perpendicular to $c$, $b$ axis and the diagonal of $ab$ plane, as well as a four-fold rotational symmetry $C_{4c}$ around $c$ axis.

We first consider the symmetry-imposed spin conductivity tensor for MnAl(001) with and without SOC (non-relativistic) as shown in Table I. The conductivity symmetries are indicated with respect to the time reversal operation, i.e., time reversal odd term ($T^{odd}$) $\tilde{\sigma}_{\mu\nu}^{s}(M) = -\tilde{\sigma}_{\mu\nu}^{s}(-M)$, and time reversal even term ($T^{even}$)

$\bar{\sigma}^s_{\mu\nu}(M) = \bar{\sigma}^s_{\mu\nu}(-M)$, where $M$ is the magnetic moment direction of MnAl ($z$ direction in this case). It is evident that without SOC, there is no off-diagonal spin Hall conductivity but a longitudinal anisotropic $\tilde{\sigma}^z_{xx}$ and $\tilde{\sigma}^z_{zz}$ with spin polarization along $z$ direction. The longitudinal spin conductivity $\tilde{\sigma}^z_{xx}$ and $\tilde{\sigma}^z_{zz}$ at 300 K are calculated to be $3.3\times10^4$ and $1.2\times10^5$ $\hbar/2e(\Omega cm)^{-1}$ by considering phonon scattering, respectively. Such large anisotropic spin conductivity is the origin of remarkable non-relativistic spin Hall effect for MnAl, as we will discuss later. Once SOC is included, both $T^{odd}$ and $T^{even}$ off-diagonal spin Hall conductivities appear, where $T^{odd}$ originates from magnetic spin Hall effect in ferromagnetic metals while $T^{even}$ comes from conventional spin Hall effect as in nonmagnetic metals Pt, W and Ta [24]. In the following, we will focus on the non-relativistic spin Hall effect and therefore ignore those SOC-driven terms.

TABLE I. The symmetry-imposed spin Hall conductivity tensor of MnAl(001) with and without SOC. The $\tilde{\sigma}$ and the $\bar{\sigma}$ correspond to the $T^{odd}$ and $T^{even}$ spin conductivity.

| Spin conductivity | Without SOC | | With SOC | |
|---|---|---|---|---|
| | $T^{odd}$ | $T^{even}$ | $T^{odd}$ | $T^{even}$ |
| $\sigma^x_{\mu\nu}$ | $\begin{pmatrix} 0 & 0 & 0 \\ 0 & 0 & 0 \\ 0 & 0 & 0 \end{pmatrix}$ | $\begin{pmatrix} 0 & 0 & 0 \\ 0 & 0 & 0 \\ 0 & 0 & 0 \end{pmatrix}$ | $\begin{pmatrix} 0 & 0 & \tilde{\sigma}^x_{xz} \\ 0 & 0 & 0 \\ \tilde{\sigma}^x_{zx} & 0 & 0 \end{pmatrix}$ | $\begin{pmatrix} 0 & 0 & 0 \\ 0 & 0 & -\bar{\sigma}^y_{xz} \\ 0 & -\bar{\sigma}^y_{zx} & 0 \end{pmatrix}$ |
| $\sigma^y_{\mu\nu}$ | $\begin{pmatrix} 0 & 0 & 0 \\ 0 & 0 & 0 \\ 0 & 0 & 0 \end{pmatrix}$ | $\begin{pmatrix} 0 & 0 & 0 \\ 0 & 0 & 0 \\ 0 & 0 & 0 \end{pmatrix}$ | $\begin{pmatrix} 0 & 0 & 0 \\ 0 & 0 & \tilde{\sigma}^x_{xz} \\ 0 & \tilde{\sigma}^x_{zx} & 0 \end{pmatrix}$ | $\begin{pmatrix} 0 & 0 & \bar{\sigma}^y_{xz} \\ 0 & 0 & 0 \\ \bar{\sigma}^y_{zx} & 0 & 0 \end{pmatrix}$ |
| $\sigma^z_{\mu\nu}$ | $\begin{pmatrix} \tilde{\sigma}^z_{xx} & 0 & 0 \\ 0 & \tilde{\sigma}^z_{xx} & 0 \\ 0 & 0 & \tilde{\sigma}^z_{zz} \end{pmatrix}$ | $\begin{pmatrix} 0 & 0 & 0 \\ 0 & 0 & 0 \\ 0 & 0 & 0 \end{pmatrix}$ | $\begin{pmatrix} \tilde{\sigma}^z_{xx} & 0 & 0 \\ 0 & \tilde{\sigma}^z_{xx} & 0 \\ 0 & 0 & \tilde{\sigma}^z_{zz} \end{pmatrix}$ | $\begin{pmatrix} 0 & -\bar{\sigma}^z_{yx} & 0 \\ \bar{\sigma}^z_{yx} & 0 & 0 \\ 0 & 0 & 0 \end{pmatrix}$ |

Although the off-diagonal spin Hall conductivity vanishes for high symmetric MnAl(001) plane, a very large spin Hall conductivity may emerge when MnAl is rotated to a lower symmetric plane. As shown in Fig. 5(a), let's suppose MnAl(001) crystal plane is rotated around $y$ axis away from $z$ axis by an angle of $\phi$, then the coordinate transformation matrix $R$ between two crystal planes is:

$$\begin{pmatrix} \cos(\phi) & 0 & -\sin(\phi) \\ 0 & 1 & 0 \\ \sin(\phi) & 0 & \cos(\phi) \end{pmatrix}. \tag{2}$$

The spin conductivity is transformed as:

$$\sigma'^{k}_{ij} = \sum_{l,m,n} R_{il} R_{jm} R_{kn} \sigma^{n}_{lm}, \tag{3}$$

where $\sigma^{n}_{lm}$ is the spin conductivity of MnAl(001), and $\sigma'^{k}_{ij}$, presented in Table II, is that of the rotated crystal plane. It is clear that now appears the off-diagonal spin Hall conductivity terms especially $\tilde{\sigma}'^{x}_{zx}$ and $\tilde{\sigma}'^{z}_{zx}$, relevant to SOT switching, are critically dependent on the anisotropic longitudinal spin conductivity of MnAl(001) as follows:

$$\tilde{\sigma}'^{x}_{zx} = -\left(\tilde{\sigma}^{z}_{xx} - \tilde{\sigma}^{z}_{zz}\right)\sin^{2}(\phi)\cos(\phi); \tilde{\sigma}'^{z}_{zx} = \left(\tilde{\sigma}^{z}_{xx} - \tilde{\sigma}^{z}_{zz}\right)\sin(\phi)\cos^{2}(\phi). \tag{4}$$

In the following, we discuss the particular case of experimentally feasible MnAl(101) plane, which corresponds to $\phi \approx 52°$. The spin current generated by non-relativistic spin Hall effect for MnAl(101) exhibits both $S_x$ and $S_z$ spin polarization, crucial for field-free switching of perpendicular magnetization.

TABLE II. Non-relativistic Spin Hall conductivity tensor of MnAl after rotating from (001) plane around $y$ axis by an angle $\phi$ away from $z$ axis.

| | $T^{odd}$ |
|---|---|
| $\sigma'^{x}_{\mu\nu}$ | $-\sin(\phi) \begin{pmatrix} \tilde{\sigma}^{z}_{xx}\cos^{2}(\phi) + \tilde{\sigma}^{z}_{zz}\sin^{2}(\phi) & 0 & \left(\tilde{\sigma}^{z}_{xx} - \tilde{\sigma}^{z}_{zz}\right)\sin(\phi)\cos(\phi) \\ 0 & \tilde{\sigma}^{z}_{xx} & 0 \\ \left(\tilde{\sigma}^{z}_{xx} - \tilde{\sigma}^{z}_{zz}\right)\sin(\phi)\cos(\phi) & 0 & \tilde{\sigma}^{z}_{xx}\sin^{2}(\phi) + \tilde{\sigma}^{z}_{zz}\cos^{2}(\phi) \end{pmatrix}$ |
| $\sigma'^{z}_{\mu\nu}$ | $\cos(\phi) \begin{pmatrix} \tilde{\sigma}^{z}_{xx}\cos^{2}(\phi) + \tilde{\sigma}^{z}_{zz}\sin^{2}(\phi) & 0 & \left(\tilde{\sigma}^{z}_{xx} - \tilde{\sigma}^{z}_{zz}\right)\sin(\phi)\cos(\phi) \\ 0 & \tilde{\sigma}^{z}_{xx} & 0 \\ \left(\tilde{\sigma}^{z}_{xx} - \tilde{\sigma}^{z}_{zz}\right)\sin(\phi)\cos(\phi) & 0 & \tilde{\sigma}^{z}_{xx}\sin^{2}(\phi) + \tilde{\sigma}^{z}_{zz}\cos^{2}(\phi) \end{pmatrix}$ |

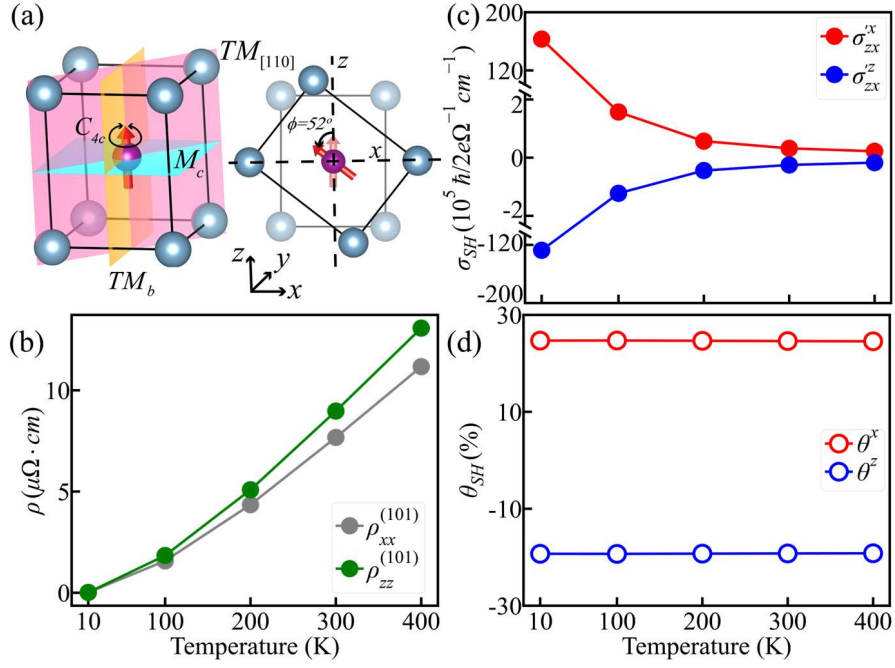

FIG. 5. (a) Atomic structure of MnAl(001) and the Cartersian coordinated system. The colored planes represent the mirror planes and $C_{4c}$ means a four-fold rotation symmetry around $c$ axis. (b) Temperature dependent longitudinal resistivity $\rho_{xx}$ and $\rho_{zz}$ for MnAl(101). (c) Spin Hall conductivities and (d) angles as a function of temperature between 10 K and 400 K for MnAl(101).

As depicted in Figs. 5(b-d), for MnAl(101) the spin Hall conductivities $\tilde{\sigma}'^{x}_{zx}$ and $\tilde{\sigma}'^{z}_{zx}$ are found to be as large as $3.2 \times 10^4$ and $-2.5 \times 10^4$ $\hbar/2e(\Omega cm)^{-1}$ at 300 K, on account of the large anisotropic transport spin polarization. There is no noticeable degradation on spin Hall conductivities even at the temperature high up to 400 K. Moreover, as shown in Fig. 5(d), the corresponding spin Hall angle $\theta^x$ and $\theta^z$ can reach to 0.25 and 0.19, much larger than the previously discussed SOC-driven spin Hall mechanism. The spin Hall angles show no obvious variation with increasing temperatures, but just slightly decrease with the temperature increasing to 400 K. This is due to the fact that the non-relativistic spin Hall conductivities $\tilde{\sigma}'^{x}_{zx}$ and $\tilde{\sigma}'^{z}_{zx}$, originating from anisotropic longitudinal spin conductivity, show the same dependence as longitudinal conductivity $\sigma_{xx}$ on band energy broadening at finite temperatures. The calculated giant spin Hall angle at room temperature indicates that the non-relativistic spin Hall effect in anisotropic ferromagnets could be highly beneficial for field-free switching.

## D. Field-free SOT switching of perpendicular magnetization

Magnetization dynamics driven by a spin current with arbitrary spin polarization can be described by the Landau-Lifshitz-Gilbert (LLG) equation as follows [12]:

$$\frac{d\hat{m}}{dt} = -\gamma \hat{m} \times \hat{B}_M + \alpha \hat{m} \times \frac{d\hat{m}}{dt} + \frac{\gamma}{M_s}[\tau_{DL}\hat{m} \times (\hat{m} \times \hat{s}) + \tau_{FL}(\hat{m} \times \hat{s})], \quad (5)$$

where $\gamma > 0$ is the gyromagnetic ratio, $\alpha$ is the damping constant, $M_s$ is the saturation magnetization and $B_M$ is an effective magnetic field. Without an external magnetic field, the effective magnetic field of a magnetic film can be expressed as $B_M = 2K_{eff}/M_s$, where $K_{eff}$ is the effective anisotropy constant. $\tau_{DL}$ and $\tau_{FL}$ represent the current induced damping-like (DL) and field-like (FL) torque per unit volume. To gain more insight into the switching of perpendicular magnetization induced by damping-like torque from spin Hall effect in 3d ferromagnets, we numerically analyze the magnetization dynamics of perpendicular magnetization by using Eq. (5), while ignoring field-like torque. By defining spin Hall ratio $\xi = \theta^z/\theta^{y,x}$ and utilizing the practical material parameters of CoFeB thin film [48], for instance $\gamma = 1.76 \times 10^{11}$ T$^{-1}$s$^{-1}$, Gilbert damping $\alpha = 0.02$, effective perpendicular anisotropy constant $K_{eff} = 1 \times 10^6$ erg/cm$^3$, saturation magnetization $M_s = 800$ emu/cm$^3$, film thickness $t_z = 1$ nm, we determine the critical torque $\tau_{DL}$ for perpendicular magnetization switching by applying a current pulse with 5 ns width. By considering $J_{sw} = (2e/\hbar)t_z\tau_{DL}/\theta_D$, where $\theta_D$ ($=\sqrt{\theta_x^2 + \theta_y^2 + \theta_z^2}$) is the magnitude of spin Hall angle, the switching current density $J_{sw}$ can also be evaluated.

Fig. 6(a) shows the switching current density ($J_{sw}$) as a function of spin Hall ratio $\xi$, for various spin Hall angles $\theta_D = 0.015, 0.02, 0.1$, and $0.25$. The $\xi$ and $\theta_D$ for particular ferromagnetic metals are indicated. It can be observed that the switching current density decreases with increasing $\xi$ when $\xi < 1$, and eventually stabilizes at the order of $10^8$ A/cm$^2$ for Fe, Co, Ni and their alloys. For ferromagnetic alloys containing heavy-element like Fe$_{50}$Pt$_{50}$, with $\theta_D \approx 0.1$ and $\xi = 0.66$ [24], the required switching current density is at the order of $10^7$ A/cm$^2$. For L1$_0$-MnAl, with giant spin Hall angle (*i.e.*, $\theta_D \approx 0.25$), exhibits the lowest switching current density among all the studied 3d ferromagnetic metals, which makes it very promising for achieving field-free SOT switching. Nevertheless, although the magnitude of switching current density in Fe, Co, Ni and their alloys is comparable to the typical values observed in non-magnetic/ferromagnetic system [10,49,50], deterministic switching of perpendicular magnetization can be achieved without an external in-plane field due to the presence of $S_z$ spin polarization. By taking $Fe_{70}Co_{30}$ as an example, Fig. 6 (b) shows the time evolution of perpendicular magnetization during the switching process. When a current pulse is applied, the perpendicular magnetization first rotates around the tilted axis and ultimately stabilizes in the -z direction after procession, thus achieving deterministic field-free switching.

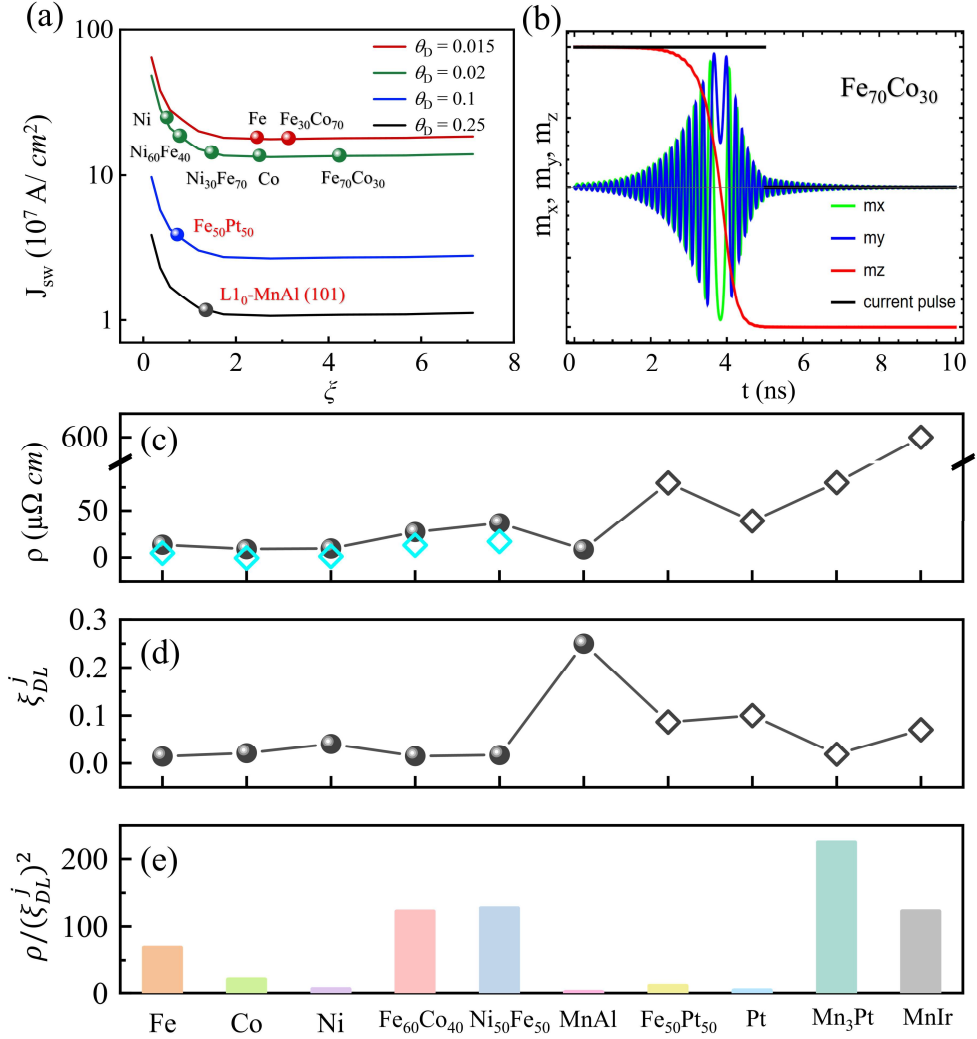

FIG. 6 (a). Switching current density ($J_{sw}$) as a function of spin Hall ratio $\xi$ ($\xi = \theta^z/\theta^{y,x}$) within spin Hall angle $\theta_D = 0.015, 0.02, 0.1$ and $0.25$. The specific ferromagnetic materials are indicated. (b) Time evolution of magnetization dynamics by using $Fe_{70}Co_{30}$ as spin current source. (c) The calculated resistivity $\rho$ compared with experimental values (cyan squares) at room temperature, along with the SOT efficiency $\xi_{DL}^j$ (d) for the considered ferromagnets (solid balls), non-magnets and antiferromagnets (black squares). (e) Comparison of the power density for 3d ferromagnets and other reported theoretical and experimental results for Pt [51], $Fe_{50}Pt_{50}$ [24], $Mn_3Pt$ [45], MnIr [12].

The switching efficiency can also be evaluated by using threshold power density $P_{sw}$, which is related to the SOT efficiency $\xi_{DL}^{j}$ (equivalent to spin Hall angle in the case of perfect interface transparency) by the expression $P_{sw} \propto \rho/(\xi_{DL}^{j})^2$ [12]. Fig. 6(c) and 6(d) show the calculated resistivity $\rho$ and SOT efficiency $\xi_{DL}^{j}$ for the studied $3d$ ferromagnets at room temperature. The reported theoretical and experimental values for Pt [51], $Fe_{50}Pt_{50}$ [24], $Mn_3Pt$ [45], and MnIr [12] are also listed for comparison. As shown in Fig. 6(c), the calculated electrical resistivity agrees well with the experimental values (cyan square) at room temperature [52-53]. Fig. 6(e) presents a comparison of threshold power density $P_{sw}$ between $3d$ ferromagnets and other reported materials. Notably, the power density of $Fe_{50}Pt_{50}$ is lower than that of $Fe_{60}Co_{40}$ and $Ni_{50}Fe_{50}$, indicating that ferromagnetic metals containing heavy-elements may be a viable strategy due to its high SOT efficiency generated by SOC-driven spin Hall effect. Furthermore, another more energy efficient approach may be utilizing non-relativistic spin Hall effect as spin current sources. For instance, $L1_0$-MnAl(101) exhibits the highest SOT efficiency and lowest power consumption among all ferromagnetic metals, making it promising for energy-efficient SOT applications.

**Summary**

In summary, we have comprehensively investigated two spin Hall mechanisms including SOC-driven spin Hall in Fe, Co, Ni and their alloys, as well as non-relativistic spin Hall mechanism in $L1_0$-MnAl. Through symmetrical analysis, we confirmed that the unconventional spin current presents in $3d$ ferromagnets, crucial for field-free perpendicular magnetization switching. Temperature and alloy composition dependencies of spin Hall effect were also examined, revealing that the spin Hall conductivities with out-of-plane spin polarization in Fe, Co, Ni and their alloys are at the order of 1000 $\hbar/2e(\Omega cm)^{-1}$ at room temperature. In $L1_0$-MnAl, the non-relativistic spin Hall angle remains large and around 0.25 at room temperature. The switching of perpendicular magnetization was analyzed by solving LLG equation, demonstrating that deterministic field-free switching can be achieved by using ferromagnetic metals as spin current sources. Notably, heavy metal doped ferromagnets may be a viable method to reduce power density, owing to their high SOT efficiency generated by SOC-driven spin Hall effect. Another more efficient approach may involve utilizing non-relativistic spin Hall mechanism, which exhibits superior energy efficiency. Our work may provide a comprehensive understanding on the spin Hall effect in ferromagnetic metals and pave the way for the design of low power consumption SOT devices.

## ACKNOWLEDGEMENT

This work was supported by open research fund of Beijing National Laboratory for Condensed Matter Physics (No. 2023BNLCMPKF010) and the National Natural Science Foundation of China (grant No. 12174129, T2394475, T2394470).

## Appendix A: Bloch spectra functions at finite temperatures

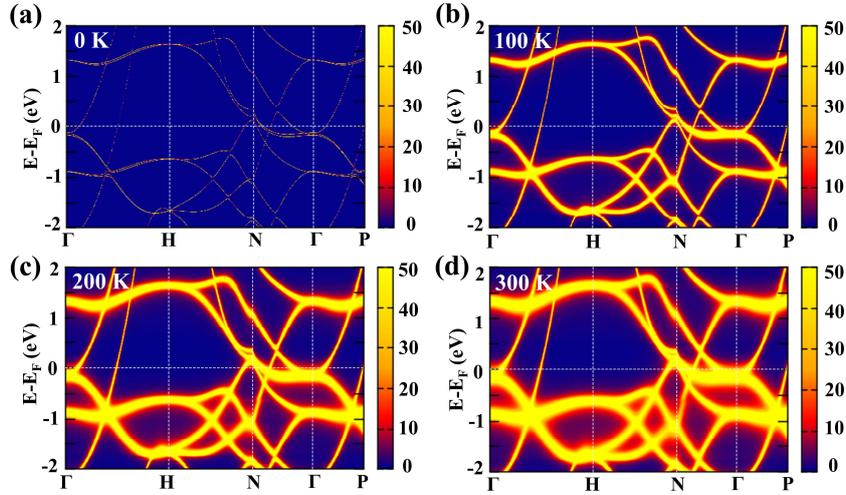

Fig. A1. The calculated Bloch spectra function (BSF) of Fe at 0 K, 100 K, 200K and 300 K.

The Bloch spectra function (BSF) at finite temperatures may qualitatively explain the transport property. Taking Fe as an example, the calculated results are shown in Fig. A1. At 0 K, the BSF of Fe exhibits zero energy broadening. As the temperature increases, electron scattering is enhanced due to phonon and spin disorder, leading to a wider band energy broadening and a shorter electron lifetime.

## Appendix B: Composition dependence of magnetic moment in $Fe_xCo_{100-x}$ and $Ni_{100-x}Fe_x$ alloy

We compare the calculated average magnetic moment with available experimental values for bulk $Fe_xCo_{100-x}$ and $Ni_{100-x}Fe_x$ alloys. As shown in Fig. A2, the calculated results agree well with the reference values. Furthermore, it can be observed that the magnetic moment in $Fe_xCo_{100-x}$ is larger than that in $Ni_{100-x}Fe_x$, indicating that the magnetization-dependent $\sigma_{zx}^z$ in $Fe_xCo_{100-x}$ may be more pronounced than that in $Ni_{100-x}Fe_x$ alloy, as shown in Figs. 3(a-b).

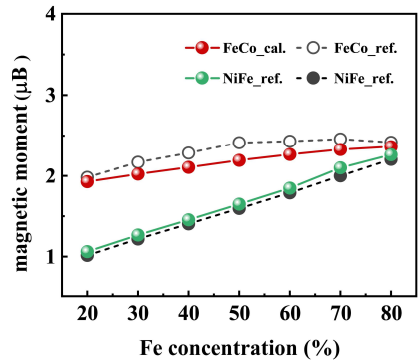

FIG. A2. The calculated average magnetic moment as a function of Fe concentration in $Fe_xCo_{100-x}$ (red balls) and $Ni_{100-x}Fe_x$ (green balls) alloys. The experimental magnetic moment of $Fe_xCo_{100-x}$ (black open circle) and $Ni_{100-x}Fe_x$ (black solid circle) are also shown for comparison [54-55].

## Reference:

[1] S. A. Wolf, D. D. Awschalom, R. A. Buhrman *et al.*, Science **294**,1488 (2001).

[2] I. Žutić, J. Fabian, and S. Das Sarma, Rev. Mod. Phys. **76**, 323 (2004).

[3] A. Manchon, H. C. Koo, J. Nitta, S. M. Frolov and R. A. Duine, Nat. Mater. **14**, 871 (2015).

[4] A. V. Larionov and L. E. Golub, Phys. Rev. B **78**, 033302 (2008).

[5] X. Qian, J. Liu, L. Fu, and J. Li, Science **346**, 1344 (2014).

[6] J. Sinova, S. O. Valenzuela, J. Wunderlich, C. H. Back, and T. Jungwirth, Rev. Mod. Phys. **87**, 1213 (2015).

[7] S.-W. Lee and K.-J. Lee, Proc. IEEE **104**, 1831 (2016).

[8] K. Ando, S. Takahashi, K. Harii, K. Sasage, J. Ieda, S. Maekawa, and E. Saitoh, Phys. Rev. Lett. **101**, 036601 (2008).

[9] L. Liu, O. -J. Lee, T. J. Gudmundsen, D. C. Ralph, and R. A. Buhrman, Phys. Rev. Lett. **109**, 096602 (2012).

[10] L. Liu, C.-F. Pai, Y. Li, H. W. Tseng, D. C. Ralph, and R. A. Buhrman, Science **336**, 555 (2012).

[11] G. Prenat, K. Jabeur, P. Vanhauwaert, *et al.*, IEEE Trans. Multi-Scale Comput. Syst. **2**, 49 (2016).

[12] A. Manchon, J. Železný, I. M. Miron, T. Jungwirth, J. Sinova, A. Thiaville, K. Garello, and P. Gambardella, Rev. Mod. Phys. **91**, 035004 (2019).

[13] X. Han, X. Wang, C. Wan, G. Yu, and X. Lv, Appl. Phys. Lett. **118**, 120502 (2021).

[14] Q. Shao, P. Li, L. Liu, *et al.*, IEEE Trans. Magn. **57**, 1 (2021).

[15] B. F. Miao, S. Y. Huang, D. Qu, and C. L. Chien, Phys. Rev. Lett. **111**, 066602 (2013).

[16] N. Nagaosa, J. Sinova, S. Onoda, A. H. MacDonald, and N. P. Ong, Rev. Mod. Phys. **82**, 1539 (2010).

[17] T. Taniguchi, J. Grollier, and M. D. Stiles, Phys. Rev. Appl. **3**, 044001 (2015).

[18] S. Iihama, T. Taniguchi, K. Yakushiji, A. Fukushima, Y. Shiota, S. Tsunegi, R. Hiramatsu, S. Yuasa, Y. Suzuki, and H. Kubota, Nat. Electron. **1**, 120 (2018).

[19] J. D. Gibbons, D. MacNeill, R. A. Buhrman, and D. C. Ralph, Phys. Rev. Appl. **9**, 064033 (2018).

[20] T. Seki, S. Iihama, T. Taniguchi, and K. Takanashi, Phys. Rev. B **100**, 144427 (2019).

[21] S. C. Baek, V. P. Amin, Y. Oh, G. Go, S.-J. Lee, M. D. Stiles, B.-G. Park, and K.-J. Lee, Nat. Mater. **17**, 509 (2018).

[22] H. Wu, S. A. Razavi, Q. Shao, X. Li, K. L. Wong, Y. Liu, G. Yin, and K. L. Wang, Phys. Rev. B **99**, 184403 (2019).

[23] T. Y. Ma, C. H. Wan, X. Wang, W. L. Yang, C. Y. Guo, C. Fang, M. K. Zhao, J. Dong, Y. Zhang, and X. F. Han, Phys. Rev. B **101**, 134417 (2020).

[24] F. Zheng, M. Zhu, J. Dong, X. Li, Y. Zhou, K. Wu, and J. Zhang, Phys. Rev. B **109**, 224401 (2024).

[25] A. Mook, R. R. Neumann, A. Johansson, J. Henk, and I. Mertig, Phys. Rev. Res. **2**, 023065 (2020).

[26] L. Salemi and P. M. Oppeneer, Phys. Rev. B **106**, 024410 (2022).

[27] M. Kimata, H. Chen, K. Kondou, S. Sugimoto, P. K. Muduli, M. Ikhlas, Y. Omori, T. Tomita, A. H. MacDonald, S. Nakatsuji, and Y. Otani, Nature **565**, 627 (2019).

[28] M. Yang, L. Sun, Y. Zeng, *et al.*, Nat. Commun. **15**, 3201 (2024).

[29] K. D. Belashchenko, Phys. Rev. B **109**, 54409 (2024).

[30] H. Ebert, D. Ködderitzsch, and J. Minár, Rep. Prog. Phys. **74**, 096501 (2011).

[31] H. Ebert et al., The Munich SPR-KKR Package, version 8.6, http://olymp.cup.uni-muenchen.de/ak/ebert/sprkkr (2017).

[32] S. H. Vosko, L. Wilk, and M. Nusair, Can. J. Phys. **58**, 1200 (1980).

[33] J. P. Perdew, K. Burke, and M. Ernzerhof, Phys. Rev. Lett. **77**, 3865 (1996).

[34] D. Ködderitzsch, K. Chadova, and H. Ebert, Phys. Rev. B **92**, 184415 (2015).

[35] K. Chadova, S. Mankovsky, J. Minár, and H. Ebert, Phys. Rev. B **95**, 125109 (2017).

[36] J. S. Faulkner and G. M. Stocks, Phys. Rev. B **21**, 3222 (1980).

[37] H. Ebert, S. Mankovsky, K. Chadova, S. Polesya, J. Minár, and D. Ködderitzsch, Phys. Rev. B **91**, 165132 (2015).

[38] D. Ködderitzsch, K. Chadova, J. Minár, and H. Ebert, New J. Phys. **15**, 053009 (2013).

[39] M. D. Kuz'min, Phys. Rev. Lett. **94**, 107204 (2005).

[40] R. S. Sundar and S. C. Deevi, Int. Mater. Rev. **50**, 157 (2005).

[41] M. Yousuf, P. Ch. Sahu, and K. G. Rajan, Phys. Rev. B **34**, 8086 (1986).

[42] K. Chadova, D. Ködderitzsch, J. Minár, H. Ebert, J. Kiss, S. W. D'Souza, L. Wollmann, C. Felser, and S. Chadov, Phys. Rev. B **93**, 195102 (2016).

[43] L. A. Hall, Survey of Electrical Resistivity Measurements on 16 Pure Metals in the Temperature Range 0 to 273 K, National Bureau of Standards, 1968.

[44] F. Wang, G. Shi, K.-W. Kim, *et al.*, Nat. Mater. **23**, 768 (2024).

[45] M. Zhu, X. Li, F. Zheng, J. Dong, Y. Zhou, K. Wu, and J. Zhang, Phys. Rev. B 110, 54420 (2024).

[46] Y. Pu, G. Shi, Q. Yang, D. Yang, F. Wang, C. Zhang, and H. Yang, Adv. Funct. Mater. **34**, 2400143 (2024).

[47] A. Jain, S. P. Ong, G. Hautier, W. Chen, W. D. Richards, S. Dacek, S. Cholia, D. Gunter, D. Skinner, G. Ceder, and K. A. Persson, APL Mater. **1**, (2013).

[48] S. Iihama, S. Mizukami, H. Naganuma, M. Oogane, Y. Ando, and T. Miyazaki, Phys. Rev. B **89**, 174416 (2014).

[49] I. Miron, K. Garello, G. Gauydin, *et al.*, Nature **476**, 189 (2011).

[50] K. Garello, I. M. Miron, C. O. Avci, F. Freimuth, Y. Mokrousov, S. Blügel, S. Auffret, O. Boulle, G. Gaudin, and P. Gambardella, Nat. Nanotechnol. **8**, 587 (2013).

[51] C. Avci, K. Garello, A. Ghosh, *et al.*, Nature Phys **11**, 570 (2015).

[52] H. Taslimi, M. Heydarzadeh Sohi, S. Mehrizi, and M. Saremi, J. Mater. Sci. Mater. Electron. **26**, 2962 (2015).

[53] G. D. Samolyuk, S. Mu, A. F. May, B. C. Sales, S. Wimmer, S. Mankovsky, H. Ebert, and G. M. Stocks, Phys. Rev. B **98**, 165141 (2018).

[54] G. Dumpich, E. F. Wassermann, V. Manns, W. Keune, S. Murayama, and Y. Miyako, J. Magn. Magn. Mater. **67**, 55 (1987).

[55] C. Jo, J. I. Lee, and Y. Jang, Chem. Mater. **17**, 2667 (2005).